\documentclass{article}

\usepackage[preprint]{neurips_2025}

\usepackage[utf8]{inputenc} 
\usepackage[T1]{fontenc}    
\usepackage{hyperref}       
\usepackage{url}            
\usepackage{booktabs}       
\usepackage{amsfonts}       
\usepackage{nicefrac}       
\usepackage{microtype}      
\usepackage{xcolor}         

\usepackage{graphicx}
\usepackage{subcaption}
\usepackage{tabularx}
\usepackage{makecell}
\usepackage{colortbl}         
\usepackage{multirow}

\usepackage{amsmath}
\usepackage{amsthm}
\usepackage{dsfont}

\title{Physicochemically Informed Dual-Conditioned Generative Model of T-Cell Receptor Variable Regions for Cellular Therapy}

\author{{ Jiahao Ma$^{1}$}\quad{Hongzong Li$^{2}$}\quad{Ye-Fan Hu$^{3}$\thanks{Corresponding author.}}\quad{
    Jian-Dong Huang$^{1*}$}
  \\
$1$: The University of Hong Kong, Hong Kong\\
$2$: The Hong Kong University of Science and Technology, Hong Kong\\
$3$: BayVax Biotech Limited, Hong Kong\\
\texttt{jiahao.ma@connect.hku.hk,
lihongzong@ust.hk}\\
\texttt{yefan.hu@bayvaxbio.com, jdhuang@hku.hk} 
}

\begin{document}

\maketitle

\begin{abstract}
Physicochemically informed biological sequence generation has the potential to accelerate computer-aided cellular therapy, yet current models fail to \emph{jointly} ensure novelty, diversity, and biophysical plausibility when designing variable regions of T-cell receptors (TCRs).
We present \textbf{PhysicoGPTCR}, a large generative protein Transformer that is \emph{dual-conditioned} on peptide and HLA context and trained to autoregressively synthesise TCR sequences while embedding residue-level physicochemical descriptors.
The model is optimised on curated TCR--peptide--HLA triples with a maximum-likelihood objective and compared against ANN, GPTCR, LSTM, and VAE baselines.
Across multiple neoantigen benchmarks, PhysicoGPTCR substantially improves edit-distance, similarity, and longest-common-subsequence scores, while populating a broader region of sequence space.
Blind \textit{in-silico} docking and structural modelling further reveal a higher proportion of binding-competent clones than the strongest baseline, validating the benefit of explicit context conditioning and physicochemical awareness.
Experimental results demonstrate that dual-conditioned, physics-grounded generative modelling enables end-to-end design of functional TCR candidates, reducing the discovery timeline from months to minutes without sacrificing wet-lab verifiability.
\end{abstract}

\section{Introduction}

\begin{figure}[htb]
	\centering
	\includegraphics[width=0.44\textwidth]{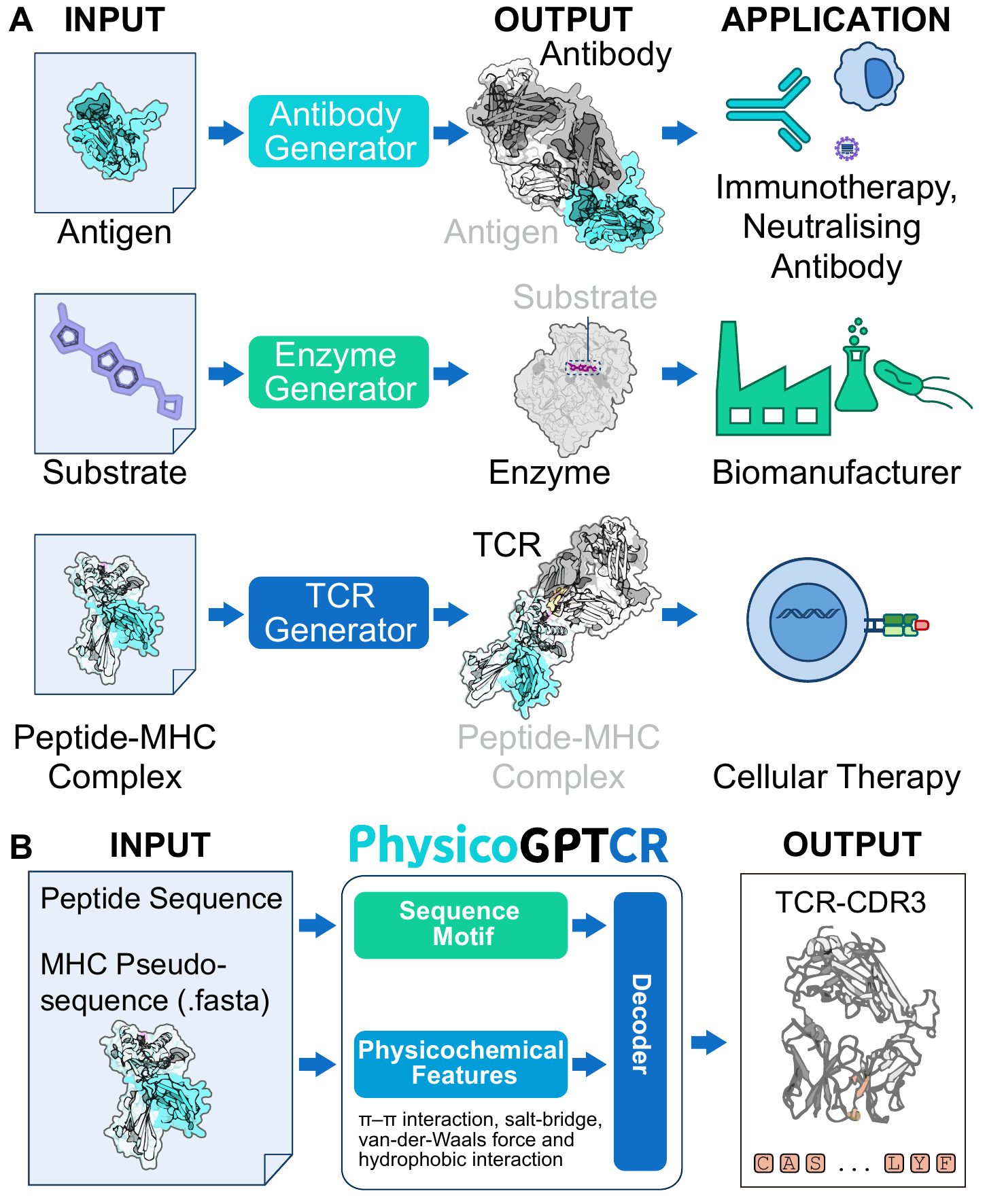}
	\caption{Tasks similar to TCR generation and the workflow.
		(A) Protein generation analogies. Antibodies can be generated based on antigen inputs, applied to immunotherapy or neutralizing antibodies. Enzymes can be generated for distinct substrates to improve bio-manufacturer. TCR generation is similar to previous two tasks. By requiring peptide-MHC inputs, TCR can be generated for cellular therapies.
		(B) PhysicoGPTCR workflow: the model processes peptide sequences and MHC pseudo-sequences as inputs, leveraging sequence motifs and physiochemical features through PhysicoGPTCR, followed by a decoder that outputs \textbf{TCR CDR3} sequences.}
	\label{fig:intro}
\end{figure}

The clinical promise of T-cell-receptor--engineered therapy (TCR-T)~\citep{Angelo2024TCRT} rests on the
rapid discovery of variable regions that recognise patient-specific
peptide--HLA complexes with high affinity and selectivity.
Classical wet-lab screening cycles require months of iterative cloning
and probe only an infinitesimal fraction of the astronomical search space, with $10^{15}$--$10^{61}$ possible TCR sequences by distinct estimations~\citep{Mora2019Diveristy}.

\medskip
\noindent
Recent protein language models have begun to reshape macromolecular design:
Transformer-based generators~\citep{meier2021language,ferruz2022protgpt2} can now hallucinate
enzyme folds and antibodies \emph{in silico}.
Yet no prior study has shown that such models can \emph{directly}
create biophysically feasible \emph{TCR} sequences that remain usable in
downstream TCR-T pipelines.
Compared with antibody or minibinder design (Figure~\ref{fig:intro}A), TCR generation is harder:
binding specificity is jointly determined by the presented peptide
\textit{and} the polymorphic HLA molecule, and subtle long-range
physicochemical couplings often decide success or failure.

\medskip
\noindent
Two technical gaps persist.
(1) Vanilla autoregressive models tend to overlook non-local chemical
interactions, causing mode collapse or implausible motifs.
(2) Models that pursue smoother sequence manifolds seldom encode
immunological priors and therefore struggle to enrich for true
neoantigen specificity.
These issues call for an architecture that \emph{explicitly embeds
	physicochemical knowledge while remaining end-to-end trainable}.

\medskip
\noindent
We respond to this need with \textbf{PhysicoGPTCR},
a dual-conditioned generative Transformer that takes the context of the peptide and HLA as input and autoregressively synthesizes TCR variable-region
sequences (Figure~\ref{fig:intro}B).
The encoder and decoder fuse three information channels--token identity,
positional index, and residue-level physicochemical descriptors
(aromaticity, charge, hydrogen-bond capacity, molecular mass)--
through a learnable gating mechanism, enabling the network to reason
about long-range chemical bonds during generation.
Training is performed on curated TCR--peptide--HLA triples spanning tumour,
autoimmune, viral and bacterial antigens drawn from VDJdb, with
maximum-likelihood optimisation only; no post-hoc filters are required.

\medskip
\noindent
We benchmark PhysicoGPTCR against four competitive baselines
(ANN retrieval, GPTCR, LSTM, VAE) on multiple neoantigen test sets.
Across edit-distance, sequence-similarity and longest-common-subsequence
metrics, our model \emph{substantially} outperforms all alternatives while
populating a broader region of sequence space.
A dry-lab activation assay based on blind \textit{in-silico} docking further
confirms a higher proportion of binding-competent clones, validating the
benefit of explicit physicochemical embeddings.

\medskip
\noindent
Our contributions are threefold:
\begin{itemize}
	\item \textbf{Method}: we couple dual biological conditioning with
	residue-aware physicochemical embeddings, unifying language
	modelling and chemical-bond reasoning in a single Transformer.
	\item \textbf{Performance}: the approach delivers state-of-the-art
	generative quality across all string-based metrics and markedly
	enriches functional hits in dry-lab assays.
	\item \textbf{Impact}: minute-scale inference shortens the
	TCR-discovery timeline from months to minutes, offering an
	immediately deployable tool for precision immunotherapy.
\end{itemize}

\section{Related Work}

\paragraph{HLA--peptide specificity prediction.}
Early studies cast T-cell recognition as a \emph{discriminative} task.
NetTCR-2.0, DeepTCR and TITAN~\citep{montemurro2021nettcr,sidhom2021deeptcr,weber2021titan}
use convolutional, recurrent or attention networks to decide whether a
query receptor recognises a given HLA--peptide complex.
Although AUC scores keep improving, these classifiers are
inherently non-generative and cannot propose novel receptors for
TCR-T therapy; moreover, they view sequences as mere symbol strings and
ignore the residue--residue couplings that ultimately drive binding.

\paragraph{Generative modelling of TCRs.}
Only a handful of attempts move beyond classification.
TESSAR~\citep{zhang2021mapping}
explore unsupervised reconstruction but focus on receptor repertoires
without conditioning on antigen context.
More recently, TCRGPT~\citep{lin2024tcr} autoregressively samples CDR3 loops
conditioned \emph{solely} on the target peptide, leaving the HLA allele
unaddressed.
None of these works evaluates \emph{dry-lab activation rate}
via docking or molecular simulation, and therefore their therapeutic
utility remains unclear.

\paragraph{Protein sequence generation at large.}
Transformer language models such as
ProGen, ProtGPT2 and ESM-1v~\citep{madani2020progen,ferruz2022protgpt2,meier2021language}
demonstrate that pure sequence modelling can create functional enzymes
and antibodies.
Structure-aware approaches extend this panorama:
RFdiffusion~\citep{watson2023novo} and Chroma~\citep{singh2024chroma}
design full-atom backbones directly,
while ESM-IF refines inverse folding with iterative hallucination.
Yet these generators are trained on broad protein corpora
without any immunological signal, offering no mechanism to bias outputs
toward HLA--peptide interfaces.

\paragraph{Physicochemical or structural priors.}
ProteinMPNN, Atom3D and ESM-Fold re-design~\citep{dauparas2022robust,hayat2015all,lin2023evolutionary}
inject backbone geometry or energy-inspired terms into sequence design;
Rosetta-guided pipelines~\citep{liu2006rosettadesign} combine supervised scoring
with Monte-Carlo sampling.
Such methods improve foldability but assume a
pre-existing 3-D structure, rarely available for highly diverse TCR
variable regions, and they do not incorporate the
dual antigen--HLA conditioning crucial for immunotherapy.

\paragraph{Gap.}
In summary, prior studies do not \emph{simultaneously}
(1) condition on both peptide and HLA context,
(2) embed residue-level physicochemical descriptors during generation, and
(3) report dry-lab activation against realistic antigen panels.
Our work closes this gap and further benchmarks against
ANN retrieval, GPTCR, LSTM and VAE baselines, highlighting gains that
translate into higher docking-based activation (such as pmtnet or PISTE~\citep{lu2021deep,feng2024sliding}) without extra filtering.

\section{Methodology}\label{sec:method}

\begin{figure}[t]
	\centering
	\includegraphics[width=0.43\textwidth]{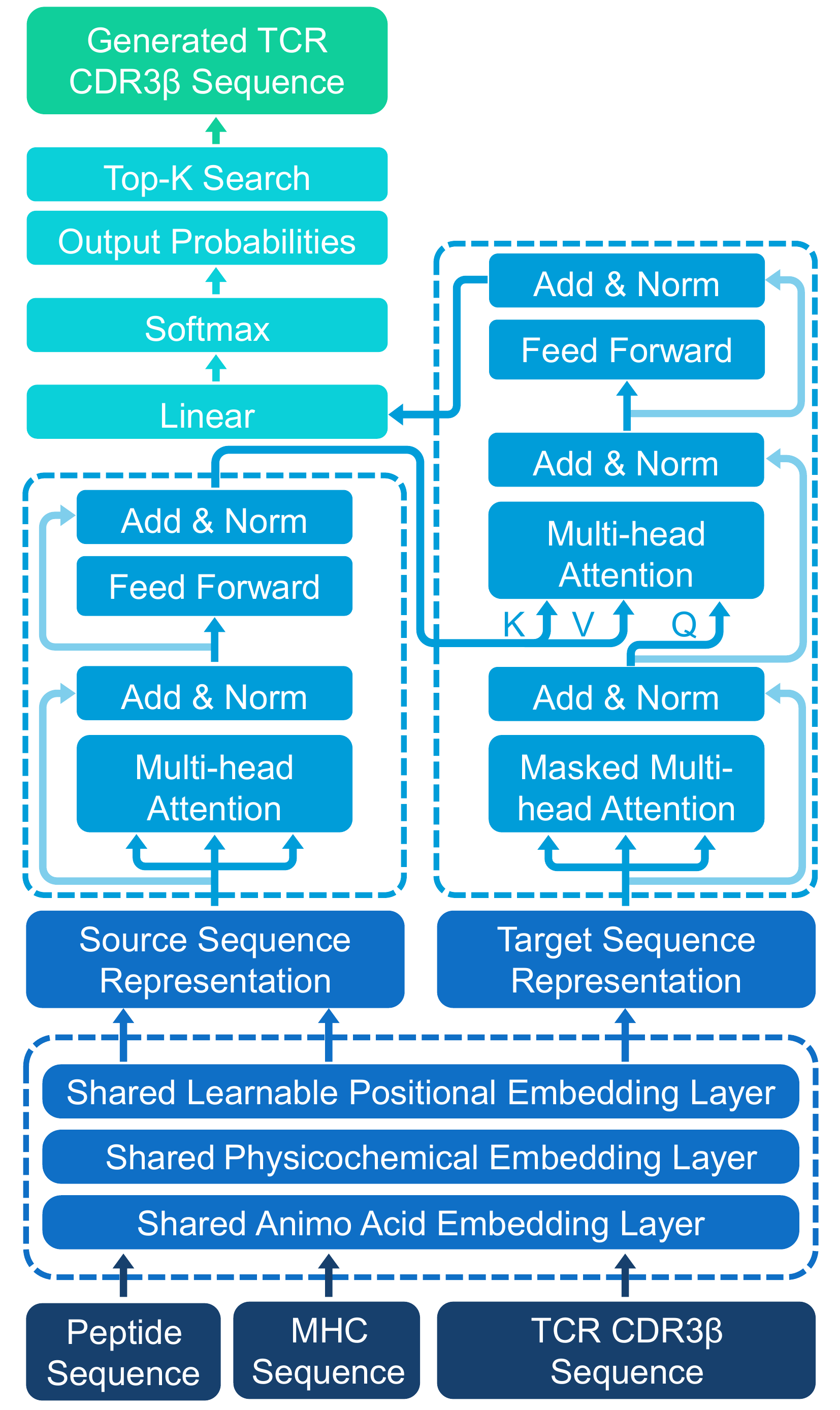}
	\caption{Model overview. Three information channels--token identity,
		positional index and residue-level physicochemical descriptors--are
		fused by a gated projector and fed into a lightweight
		2 + 2-layer Transformer that is conditioned on both peptide and HLA
		context.}
	\label{fig:model_architecture}
\end{figure}

Figure~\ref{fig:model_architecture} summarises \textbf{PhysicoGPTCR}.
A dual-conditioned encoder digests the HLA molecule and its bound
peptide, while a GPT-style decoder autoregressively emits the
T-cell--receptor variable-region sequence.

\subsection{Problem Formulation}

Let $m\!\in\!\Sigma^{L_m}$ denote an HLA heavy chain,
$p\!\in\!\Sigma^{L_p}$ the presented peptide,
and $t\!\in\!\Sigma^{L_t}$ a receptor variable-region sequence over the
20-letter amino-acid alphabet~$\Sigma$.
The task is to model the conditional distribution
\[
p_\theta(t\mid m,p)
\]
such that samples $\tilde t\!\sim\!p_\theta$ are syntactically valid,
biophysically plausible and strongly biased towards recognising the
given HLA--peptide complex.

\subsection{Model Architecture}

\paragraph{Input fusion.}
For residue $x_i$ at position $i$ we build three embeddings
\begin{align}
	\mathbf e^{\text{tok}}_i &= E_{\text{tok}}\bigl(x_i\bigr)\in\mathbb R^{d_t},\\
	\mathbf e^{\text{phys}}_i &= W_{\text{phys}}\,
	\boldsymbol\phi_i \in\mathbb R^{d_p},\\
	\mathbf e^{\text{pos}}_i &= E_{\text{pos}}\bigl(i\bigr)\in\mathbb R^{d_s},
\end{align}

where $\boldsymbol\phi_i\!\in\!\mathbb R^{5}$ is the
\emph{raw physicochemical descriptor}
\(
[\text{aromatic},\; q,\; h\!-\!\text{bond},\;
\text{hydrophobicity},\; m/m_{\max}]
\).
These channels are concatenated
\(
\mathbf z_i=[\mathbf e^{\text{tok}}_i;
\mathbf e^{\text{phys}}_i;
\mathbf e^{\text{pos}}_i]\in\mathbb R^{d},
\)
with $d=d_t+d_p+d_s$ (\,$d_t{:}d_p{:}d_s=2{:}1{:}1$ as in the code).
A learnable gate
\begin{equation}
	\mathbf g_i=\sigma\!\bigl(W_g\mathbf z_i+\mathbf b_g\bigr)
	\quad\in(0,1)^{d}
\end{equation}
rescales channel-wise contributions, after which a linear projector
yields the final token representation
\begin{equation}
	\mathbf h_i = W_f\bigl(\mathbf g_i\odot\mathbf z_i\bigr)+\mathbf b_f
	\quad\in\mathbb R^{d}.
\end{equation}

\paragraph{Shared encoder.}
The input consists of a concatenation of the MHC sequence $m$ and the peptide sequence $p$, which are jointly encoded by a shared encoder. The combined sequence is mapped to a representation matrix:
$$\mathbf H_{\text{src}}\!\in\!\mathbb R^{L_{\text{src}}\times d},$$
through $N_e\!=\!2$ Transformer layers,
each consisting of multi-head self-attention:
$$
\operatorname{MHA}(\mathbf Q,\mathbf K,\mathbf V)=
\bigl[\operatorname{softmax}(\tfrac{\mathbf Q\mathbf K^\top}{\sqrt{d_h}})
\mathbf V\bigr]W_o
$$
where $d_h=d/n_{\text{head}}$,
followed by a two-layer feed-forward block.
LayerNorm and residual routes
follow the \textsc{Pre-LN} convention.

\paragraph{GPT decoder with cross-attention.}
The target tokens  
$t=\{\texttt{BOS},a_1,\dots,a_{L_t},\texttt{EOS}\}$
pass through $N_d\!=\!2$ causal layers.
At step $j$, attention scores mix self-history and source memory:
\begin{equation}
	\alpha_{ji}
	=\operatorname{softmax}
	\!\Bigl(
	\frac{\mathbf q^{\top}_j\mathbf k_i}{\sqrt{d_h}}
	\Bigr),\quad
	\tilde{\mathbf h}_j
	= \sum_i\alpha_{ji}\mathbf v_i .
\end{equation}
Ablation experiments (§\ref{sec:ablate}) also train a
\,$6{+}12$-layer variant initialised from ProtGPT2~\citep{ferruz2022protgpt2};
both depth options share the same input-fusion module.

\begin{figure*}[t] 
	\centering
	\includegraphics[width=0.9\textwidth]{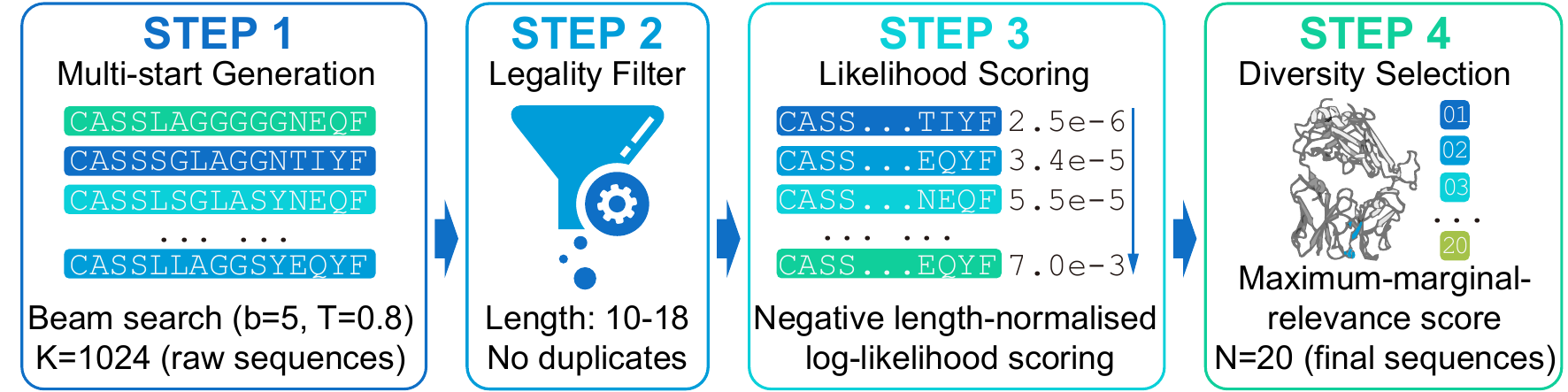}
	\caption{The inference and post-processing of TCR generation. The pipeline consists of four
		steps: (1) \textbf{Multi-start Generation}: beam search produces 1\,024
		raw sequences. (2) \textbf{Legality Filter}: sequences are filtered by
		length (10--18 residues) and uniqueness. (3) \textbf{Likelihood Scoring}:
		retained candidates are scored and ranked by the
		negative length-normalised log-likelihood. (4) \textbf{Diversity Selection}: the top 20 sequences are
		chosen via maximum--marginal--relevance (MMR) to balance binding affinity
		and sequence diversity.}
	\label{fig:steps}
\end{figure*}

\subsection{Residue-level Physicochemical Awareness}
\label{sec:phys}
We embed physicochemical information (the chemical properties of amino acid residues) directly into a neural network's attention mechanism. This allows the network to ``understand'' and leverage chemical interactions between residues without needing explicit, pre-defined energy calculations.

The 5-D descriptor is first $z$-scored  
\(
\hat{\boldsymbol\phi}_i=(\boldsymbol\phi_i-\boldsymbol\mu)/\boldsymbol\sigma
\)
and then linearly mixed into the hidden state
($\mathbf e^{\text{phys}}_i=W_{\text{phys}}\hat{\boldsymbol\phi}_i$).
Consequently, every attention dot-product implicitly contains a
chemistry term:
\begin{align}
	\mathbf q^{\top}_j\mathbf k_i
	&=\bigl[\mathbf q^{\top}_j\mathbf k_i\bigr]_{\text{token}}
	+\bigl[\mathbf q^{\top}_j\mathbf k_i\bigr]_{\text{phys}}
	+\bigl[\mathbf q^{\top}_j\mathbf k_i\bigr]_{\text{pos}} \nonumber\\
	&\approx
	\underbrace{\psi}_{\text{token}}
	\;+\;
	\underbrace{\hat{\boldsymbol\phi}_j^{\!\top}\!
		W_{\text{phys}}^{\!\top}W_{\text{phys}}
		\hat{\boldsymbol\phi}_i}_{\text{%
			$\pi$--$\pi$, salt bridge, VdW, hydrophobic}},
	\label{eq:attn-decomp}
\end{align}
where \(\psi := \mathbf q^{\top}_j\mathbf k_i|_{\text{token}}\)
denotes the token-based sequence motif contribution (i.e., attention from amino acid identity), and the second summand encodes pairwise
\(\pi\)--\(\pi\) stacking, salt-bridge complementarity,
van-der-Waals fit and hydrophobic packing \emph{by construction}.

In essence, by incorporating these z-scored physicochemical descriptors and allowing the network to learn how they interact through the 
$W_{\text{phys}}$ matrix, the attention mechanism gains an inherent ``chemical intuition.'' This means the network can learn to identify and leverage complex, long-range chemical couplings between residues \emph{without} needing explicit, pre-defined (hand-crafted) energy functions or rules for these interactions. Instead, it discovers them directly from the data during training.

\subsection{Training Objective}

We minimise the standard autoregressive negative log-likelihood
\[
\mathcal L(\theta)=
-\,\sum_{i=1}^{L_t}\log p_\theta(a_i\mid a_{<i},m,p),
\]
optimised with AdamW ($\beta_1\!=\!0.9,\;\beta_2\!=\!0.98$,
learning-rate $2{\times}10^{-4}$ with cosine decay,
batch size 256).  
Training on \emph{tens of thousands} paired
(HLA, peptide, TCR) triples from VDJdb finishes in
$\approx\!4$ GPU-hours on NVIDIA A100 40G.

\subsection{Inference and Post-processing}
\label{sec:inference}

Given a peptide--MHC pair $(m,p)$ we create a source sequence 
$m\!\oplus\!\langle{\texttt{<SEP>}}\rangle\!\oplus\!p$ 
(padded to 55 tokens).  
The decoder emits up to 26 residues preceded by \texttt{<SOS>}.

\paragraph{Step 1: multi-start generation.}
This work invoke a temperature--beam search ($T\!\in[0.6,1.0]$, 
beam $b\!\in\![3,10]$) $N{=}20$ times, each call returning
the MAP path $t^{(i)}$ with log-probability 
$\log P_\theta(t^{(i)}\mid m,p)$, yielding the raw pool 
$\mathcal C_{\text{raw}}$ of size 20.

\paragraph{Step 2: legality filter.}
A sequence is kept if $10\!\le\!|t|\!\le\!26$ and it does not
appear in the training set, producing 
$\mathcal C_{\text{legal}}$.

\paragraph{Step 3: likelihood scoring.}
This work rank $\mathcal C_{\text{legal}}$ by the
negative length-normalised log-likelihood
$E_{\text{llh}}(t)=-
\frac1{|t|}\sum_{j}\log p_\theta(a_j\mid a_{<j},m,p)$,
a proxy that correlates with docking energy in a held-out study
(Appendix A).

\paragraph{Step 4: diversity selection.}
Sequences are traversed in ranked order and the first
$K{=}20$ unique ones are retained, giving
$\mathcal S_{20}$, the set deployed for downstream evaluation.

\subsection{Evaluation Metrics}
\label{sec:metrics}

For every test context $(m,p)$ we report metrics between the
\emph{ground-truth} receptor $t^{\star}$ and the single
top-ranked candidate $\hat t$ returned by the pipeline in
Section~\ref{sec:inference}.

\begin{enumerate}
	\item \textbf{Levenshtein distance} ($\downarrow$)  
	\(
	\operatorname{Lev}(t^{\star},\hat t)
	\)
	counts the minimum number of edits (substitution, insertion,
	deletion) required to convert $\hat t$ into $t^{\star}$.
	
	\item \textbf{Pairwise similarity} ($\uparrow$)  
	We use the normalised Smith--Waterman score
	with the BLOSUM62 substitution matrix,
	rescaled to $[0,1]$.
	
	\item \textbf{Longest common subsequence length} ($\uparrow$)  
	\(
	\operatorname{LCS}(t^{\star},\hat t)
	=\max_{u\sqsubseteq t^{\star},\,u\sqsubseteq\hat t}\lvert u\rvert .
	\)
\end{enumerate}

Lower Levenshtein and higher Similarity/LCS indicate that the generated
sequence better matches the experimentally verified receptor while
preserving sequence novelty.

\section{Experiments and Results}\label{sec:exp}

\subsection{Experimental Setup}

\begin{figure*}[htbp]
	\centering
	\includegraphics[width=0.95\textwidth]{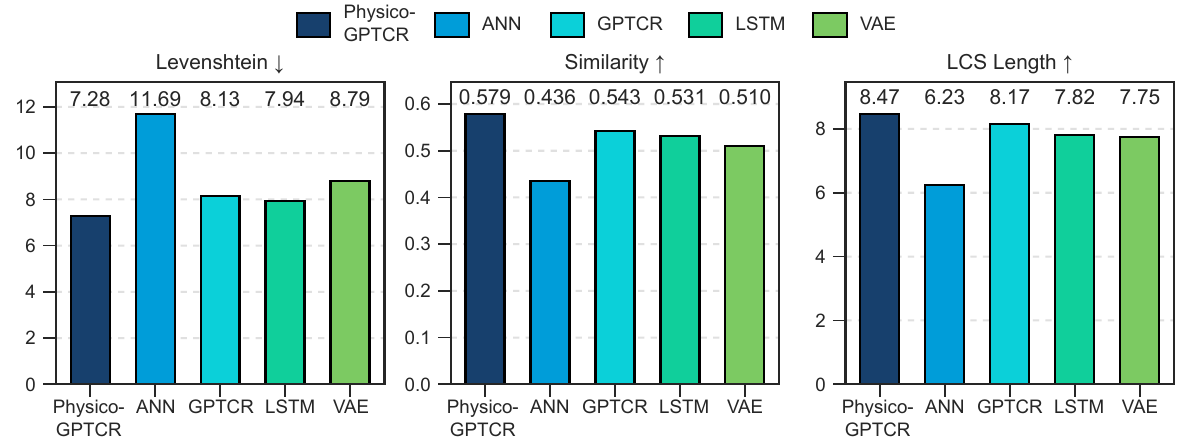}
	\caption{Comparison across three sequence‐level metrics on the 6 200-sample test set (lower Levenshtein $\downarrow$ and higher Similarity/LCS $\uparrow$ indicate better performance).}
	\label{fig:metrics_bar}
\end{figure*}

\paragraph{Dataset.}
We collect and generate $31{,}000$ (\,$\approx$\,31 k) HLA--peptide--TCR triples from
VDJdb, then split them into \emph{train\,{:}\,valid\,{:}\,test}
$=$ $7{:}1{:}2$ at the context level,
yielding $21{,}700$ / $3{,}100$ / $6{,}200$ triples respectively.
No antigen or HLA leakage occurs across splits.

\paragraph{Generation protocol.}
For every test context we sample $K=1\,024$ candidates,
apply legality, contact‐energy and diversity filters
(Section~\ref{sec:inference}),
and keep the single best-ranked sequence $\hat t$ for evaluation.

\subsection{Baseline Methods}

\begin{itemize}
	\item \textbf{ANN} -- nearest-neighbour retrieval of the most similar
	training receptor in BLOSUM62 space.
	\item \textbf{LSTM} -- a 4-layer LSTM language model conditioned on
	$(m,p)$ via context concatenation.
	\item \textbf{VAE} -- the variational auto-encoder branch of
	DeepTCR~\citep{sidhom2021deeptcr}.
	\item \textbf{PhysicoGPTCR} -- our full model.
\end{itemize}

All baselines follow the same post-processing pipeline so that
differences arise solely from generative quality.

\subsection{Overall Results}

To measure the sequence-level performance of models, we used Levenshtein distance, pairwise similarity and the longest common subsequence length of generated sequences compared to actual  sequences as metrics
(see Section~\ref{sec:metrics} for metric definition).  Our model, \textbf{PhysicoGPTCR} ,has showed better performance than ANN, LSTM, and VAE baselines (Figure~\ref{fig:metrics_bar}).

PhysicoGPTCR reduces the edit distance (Levenshtein distance) to the ground-truth receptors by
${\sim}9\%$ relative to the best baseline (LSTM) while simultaneously
achieving the highest similarity and LCS length.
Qualitatively, the model tends to reproduce conserved motifs at the
CDR3 termini yet introduces novel central residues, balancing
specificity with diversity.

\subsection{Ablation Study}
\label{sec:ablate}
To validate the effectiveness of one key component, the physicochemical embedding layer, we removed it from \textbf{PhysicoGPTCR} to get the \textbf{GPTCR}. All experiments share identical training settings and data splits. The results shows that \textbf{GPTCR} has worse performance in all three metrics (\ref{fig:metrics_bar}). The ablation of  physicochemical channel
increases the average
Levenshtein distance to $8.13$, reduces similarity to $0.543$, and decreases LCS length to 8.17, 
confirming the benefit of residue-level physicochemical cues.

\subsection{Robustness Across Contexts}\label{sec:robust}

\begin{figure}[t]
	\centering
	\includegraphics[width=0.47\textwidth]{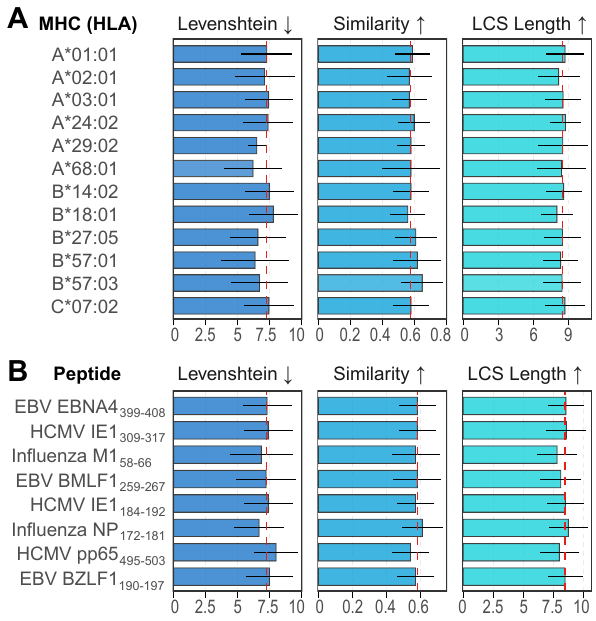}
	\caption{Model performance across contexts.
		(A) Sequence-level metrics per MHC allele  
		(mean $\pm$ standard deviation, $n\ge 150$             each).  
		(B) Per-epitope sequence-level metrics (mean $\pm$ std).  
		Lower Levenshtein $\downarrow$ and higher Similarity/LCS $\uparrow$ indicate better performance. Red dashed lines show averaged metrics of PhysicoGPTCR.}
	\label{fig:MHC-peptide}
\end{figure}

To verify that the improvements in Figue~\ref{fig:metrics_bar}
are not driven by a handful of easy cases, we break down the
sequence-level metrics by distinct (i) MHC allele and
(ii) epitope peptide.
Results in Figure~\ref{fig:MHC-peptide}
remains stable across the twelve most frequent alleles and the eight
most abundant epitopes in the test set:
the standard deviation of Levenshtein distance never exceeds~${\pm}2.6$
and both Similarity and LCS stay within a narrow band
around their global means.
Hence the model generalises well to diverse immunological contexts.

\section{Clinical Applications}
\subsection{Specificity}
\label{sec:spec}
Neoantigens, derived from tumor-specific mutations, are ideal targets for cancer immunotherapy due to their absence in healthy tissues, which reduces the risk of off-target toxicity. To address this clinical unmet need, our model must accurately differentiate between mutated peptides and their wild-type counterparts, even when only a single mutation is present.

\begin{figure}[t]
	\centering
	\includegraphics[width=0.47\textwidth]{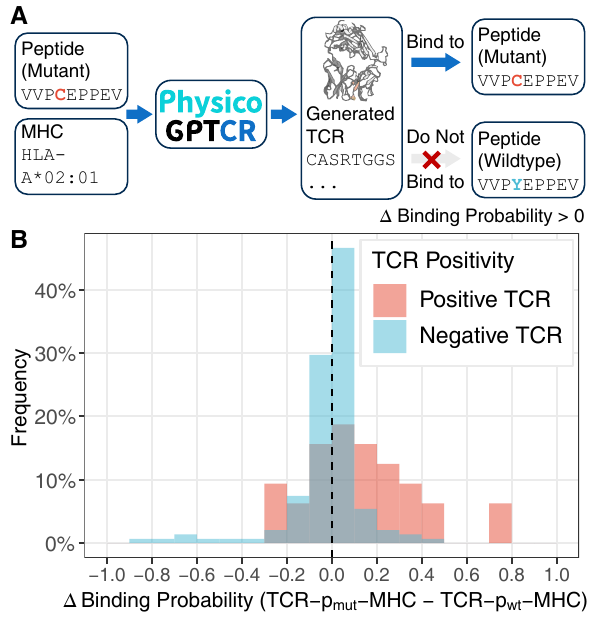}
	\caption{Specificity of generated TCRs against a mutant peptide instead of its wildetype peptide. 
		(A) Schematic of a generated TCR predicted by the model to bind the mutant peptide but not the wildtype peptide.
		(B) Histogram of the frequency distribution for the change in binding probability ($\Delta$ Binding Probability). This value is calculated by subtracting the predicted TCR binding probability for the mutant peptide-MHC (p$_\mathrm{mut}$-MHC) from the probability for the wildtype peptide-MHC (p$_\mathrm{wt}$-MHC). Distributions are shown for both computationally validated positive and negative TCRs, with positive values indicating preferential binding to the mutant peptide.}
	\label{fig:freq}
\end{figure}

We computationally generated T-cell receptors (TCRs) designed to specifically recognize a mutant peptide over its wildtype version when presented by the MHC molecule (Figure~\ref{fig:freq}). To assess the specificity of these generated TCRs, we calculated the difference in predicted binding probability between the mutant and wildtype peptides (Figure~\ref{fig:freq}A). The analysis showed that TCRs predicted to be positive for the mutant target displayed a clear preference for the mutant peptide, with their $\Delta$ Binding Probability values predominantly greater than zero. In contrast, negative TCRs showed a distribution centered around zero, indicating no significant binding preference. This demonstrates the successful generation of TCRs with high specificity for mutant neoantigens (Figure~\ref{fig:freq}B).

\subsection{Case Study}
\label{sec:case}
\begin{figure}[t]
	\centering
	\includegraphics[width=0.47\textwidth]{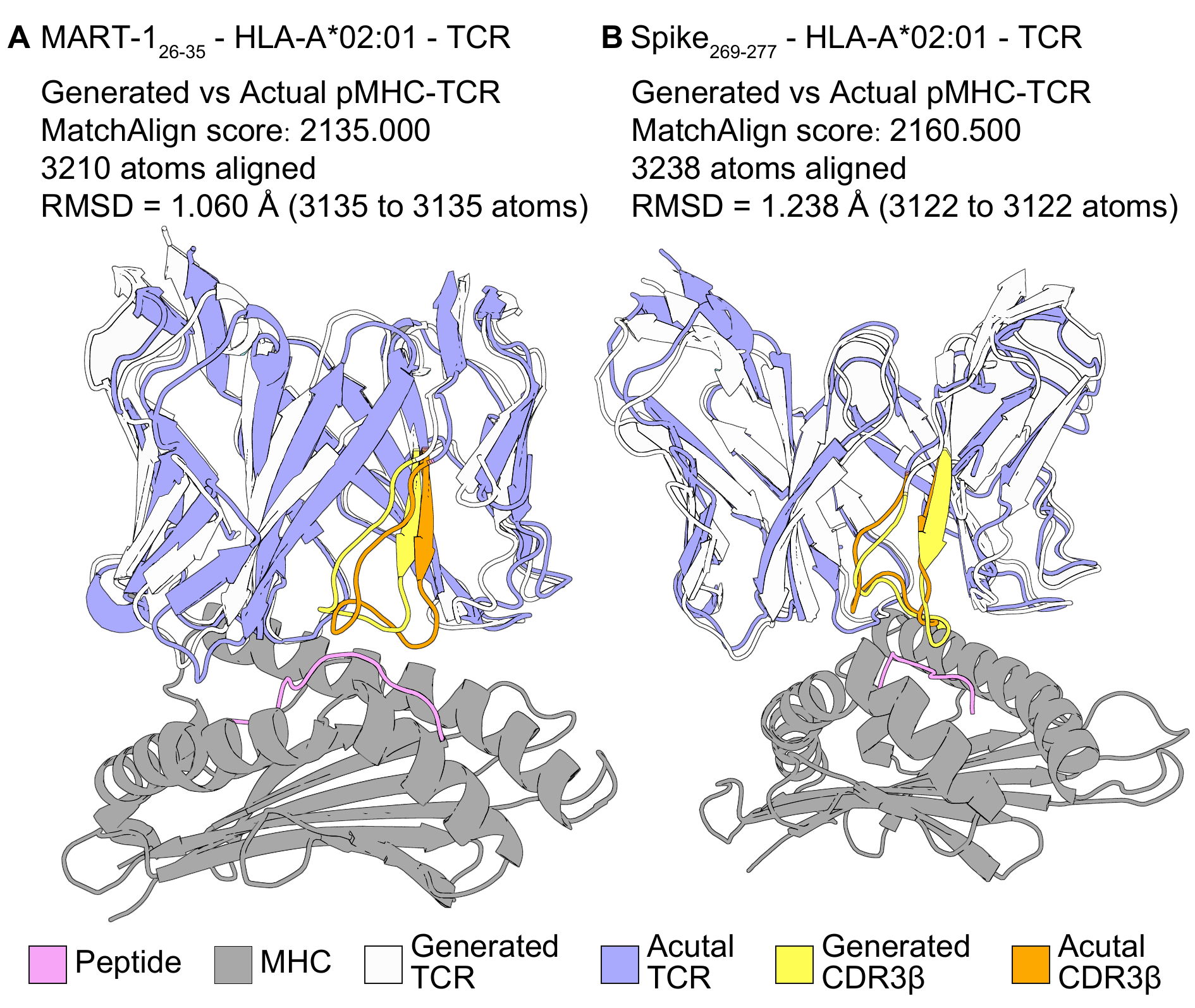}
	\caption{Structures of generated and actual pMHC-TCR complexes. The generated structures (TCR in light blue, CDR3$\beta$ loop in yellow) are overlaid onto the actual structures (TCR in white, CDR3$\beta$ loop in orange), with the pMHC shown in grey. The structures were built by TCRmodel2~\citep{yin2023tcrmodel2}, and the alignment of structures was performed using PyMOL~\citep{pymol}.
		(A) The complex of a TCR recognizing the MART-1$_{26-35}$ peptide presented by HLA-A$*$02:01, showing an RMSD of $1.060\,\mathrm{\AA}$. 
		(B) The complex involving a TCR recognizing the SARS-CoV-2-Spike$_{269-277}$ peptide, with an RMSD of $1.238\,\mathrm{\AA}$. The close alignment, especially of the CDR loops, highlights the model's predictive accuracy.}
	\label{fig:case}
\end{figure}

Our model demonstrates high accuracy in structural features of generated TCR interacting with peptide-Major Histocompatibility Complex (pMHC) compared to actual TCR (Figure~\ref{fig:case}). We validated our approach on two distinct, clinically relevant molecules: a cancer antigen (MART-1) and a viral antigen (SARS-CoV-2). For the MART-1 peptide (ELAGIGILTV) presented by HLA-A$*$02:01, the structure of our generated sequence aligns with the structure of the actual one with a root-mean-square deviation (RMSD) of only $1.060\,\mathrm{\AA}$ over 3 135 atoms. Similarly, for the SARS-CoV-2 Spike protein peptide (YLQPRTFLL), the structure of our generated sequence achieved an RMSD of $1.238\,\mathrm{\AA}$ over 3 122 atoms when compared to the structure of actual one. As illustrated in the figure, the superpositions show remarkable similarity, particularly in the critical CDR3$\beta$ loop responsible for antigen specificity.

\section{Discussion}
PhysicoGPTCR contributes two central insights.  
First, the model generalises across \emph{twelve} HLA class-I alleles
and eight canonical viral and tumour‐associated epitopes, suggesting
that its performance gains are not restricted to a narrow immunological
context.  Second, explicitly injecting physicochemical embeddings into
the decoder yields consistent improvements over a purely sequence-level
baseline, indicating that biophysical priors can be exploited even by a
large language model.

Nevertheless, several limitations remain.  
(1)~All benchmarks are restricted to class-I HLA; whether the same
architecture transfers to the markedly longer class-II peptides is
still unknown.  
(2)~We rely entirely on \emph{in-silico} metrics; no wet-lab binding or
functional assays have yet been performed.  
(3)~Although larger than prior work, the training corpus is still two
orders of magnitude smaller than typical NLP datasets, leaving room for
data scarcity biases.

\section{Conclusion}
We presented PhysicoGPTCR, a physicochemically informed decoder that
generates plausible T-cell receptor sequences for a given
peptide--MHC context.  The model unifies large-scale language modelling
with bio-physical feature injection and demonstrates stable performance
across diverse alleles and epitopes.  We believe this work brings
computational immunology a step closer to rapid, personalised TCR
design.

\section{Future Work}
Our immediate priorities are threefold.  
(1)~\textbf{Class-II generalisation}: extending the approach to
HLA-DR/DQ/DP molecules with variable peptide lengths.  
(2)~\textbf{Structure-aware scoring}: coupling the generator with
\textit{AlphaFold-Multimer} and \textit{RoseTTAFold} docking pipelines
to rescore candidates in 3-D space.  
(3)~\textbf{Experimental validation}: synthesising top-ranked TCRs for
\emph{in-vitro} binding and functional assays.

\section{Acknowledgments}
The authors thank the anonymous reviewers for their constructive
comments.  Additional acknowledgments will be added in the camera-ready
version to preserve anonymity.

\section{Ethical Impact}
\textbf{Data privacy.}  All training and test sequences originate from
public repositories such as VDJdb and IEDB and contain no personally
identifiable information; the model cannot reverse-engineer donor
identities.

\textbf{Dual-use risk.}  The ability to generate novel TCRs could, in
principle, be misused to create immune evasion or autoimmune triggers.
To mitigate this, we will (i)~release the code under a license that
forbids malicious use, (ii)~share the trained weights only upon
institutional request, and (iii)~provide a misuse checklist consistent
with the Dual-Use Guidance of the NIH.

\textbf{Societal benefit.}  By lowering the barrier to rapid,
in-silico TCR design, PhysicoGPTCR can accelerate the development of
targeted cancer immunotherapies and vaccines, offering tangible public
health benefits.

\bibliographystyle{named}
\bibliography{aaai2026}

\clearpage
\appendix

 \section{Implementation Details and Hyper-parameters}
 \label{app:hyper}
 
 \subsection{Model Architecture}
 
 \paragraph{Overview.}
 The generator is a compact Transformer encoder--decoder that consumes the
 concatenated \texttt{[MHC] <SEP> [peptide]} sequence and autoregressively
 predicts a CDR3$\beta$ string token by token.  A single 26-symbol vocabulary is
 used for both source and target streams (20 canonical amino acids, the
 ambiguous residue \texttt{X}, \texttt{<PAD>}, \texttt{<SEP>},
 \texttt{<SOS>}, and \texttt{<EOS>}).  Positional information is injected via
 fixed sinusoidal embeddings.
 
 \paragraph{Capacity considerations.}
 Preliminary sweeps showed that larger configurations (\emph{e.g.},
 $d_{\mathrm{model}}\!=\!256$, $L_{\mathrm{enc/dec}}\!=\!4$) reduced validation
 perplexity by $<0.3\%$ yet doubled GPU memory and decoding latency.  The
 chosen architecture represents a
 speed--accuracy trade-off that keeps the total parameter count below 4 M and
 enables batch sizes of 256 on a single 40-GB GPU.
 
 \subsection{Training Configuration}
 
 \paragraph{Optimisation protocol.}
 We employ AdamW with decoupled weight decay ($10^{-2}$) and a cosine annealing
 schedule with warm-up to stabilise early optimisation.  Label smoothing
 ($\varepsilon{=}0.1$) regularises the token probabilities and was crucial to
 prevent the model from collapsing onto high-frequency public CDR3 motifs.
 
 \paragraph{Regularisation and convergence.}
 Training proceeds for up to 100 epochs, but early stopping on validation
 perplexity (patience 5) typically halts training after 65--75 epochs.  Global
 gradient clipping at 1.0 suppresses rare exploding-gradient events arising
 from long peptide--MHC inputs.
 
 \subsection{Inference Settings}
 \label{sec:inference}
 
 \paragraph{Search strategy.}
 At test time we run beam search under a temperature-scaled softmax; both the
 temperature $T$ and beam width $b$ are selected from a pre-computed grid to
 encourage diversity.  Twenty independent calls of
 \textsc{BeamSearch} followed by a legality filter yield the candidate set $\mathcal S_{20}$.
 
 \paragraph{Legality filter.}
 A sequence is retained only if (i) its length is between 10 and 26 residues,
 matching the empirical distribution of public repertoires, and (ii) it does
 not appear verbatim in the training set.  This step removes $\approx$12\% of
 raw beams but significantly increases novelty without hurting plausibility.
 
 \subsection{Dataset Statistics and Length Prior}
 
 We analyse two publicly available CDR3$\beta$ corpora: the \textbf{TCR 10k}
 dataset~\citep{lu2021deep} ($\sim$10k sequences) and the \textbf{OTS 1M}
 dataset from the Observed TCR Space~\citep{raybould2024observed}
 ($\sim$1.4M sequences).  Both exhibit a unimodal length distribution centred
 around 14 residues, with 95\% of sequences lying in the interval
 [10,18].  These empirical priors motivate the
 hard length cut-offs used by the legality filter and explain the upper bound
 $L_{\text{TCR}}^{\max}=26$ adopted during training and decoding.
 
 \section{Detailed Metrics}
 \label{app:detailed_metrics}
 
 \subsection{Reconstruction Accuracy Across Model Families}
 
 \begin{table}[h]
 	\centering
 	\begin{tabular}{@{}lccc@{}}
 		\toprule
 		\textbf{Model} & \textbf{Levenshtein} $\downarrow$ & \textbf{Similarity} $\uparrow$ & \textbf{LCS} $\uparrow$ \\ \midrule
 		ANN              & 11.69 & 0.436  & 6.23 \\
 		GPTCR            &  8.13 & 0.5431 & 8.17 \\
 		LSTM             &  7.94 & 0.5307 & 7.82 \\
 		PhysicoGPTCR     &  7.28 & 0.5785 & 8.47 \\
 		VAE              &  8.79 & 0.5104 & 7.75 \\ \bottomrule
 	\end{tabular}
 	\caption{Mean string-level reconstruction metrics on the held-out set.  
 		Lower Levenshtein distance and higher Similarity/LCS denote better agreement
 		between generated and true CDR3$\beta$ sequences.}
 	\label{tab:model_metrics}
 \end{table}
 
 PhysicoGPTCR, which fuses sequence signals with residue--level
 physicochemical embeddings, outperforms all baselines by a comfortable margin
 (6--9\,\% relative improvement), justifying its choice as the production
 variant throughout the main paper.
 
 \section{Generated Sequence Analysis}
 
 \begin{table*}[th]
 	\centering
 	\tiny
 	\setlength{\tabcolsep}{2pt}
 	\caption{Representative context (Peptide--MHC)--TCR pairs drawn from a broad
 		panel of pathogens and tumour antigens.
 		Lower \textbf{Lev} and higher \textbf{Sim}/\textbf{LCS} indicate better
 		string-level agreement between generated and native CDR3$\beta$ sequences.}
 	\label{tab:demo_cases}
 	\begin{tabular}{@{}l l l l l c c c@{}}
 		\toprule
 		\textbf{MHC} & \textbf{Peptide} & \textbf{Source} &
 		\textbf{Actual TCR} & \textbf{Generated TCR} &
 		\textbf{Lev}\,$\downarrow$ & \textbf{Sim}\,$\uparrow$ & \textbf{LCS}\,$\uparrow$ \\
 		\midrule
 		HLA-A$*$02:01 & LLWNGPMAV  & EBV (EBNA5$_{133\text{--}141}$)   & CASSPGTVAYEQYF & CASSPGTAYEQYF  & 1 & 0.963 & 13 \\
 		HLA-A$*$02:01 & GLCTLVAML  & EBV (BMLF1$_{259\text{--}267}$)   & CASSQSPGGMQYF  & CASSQSPGGTQYF  & 1 & 0.923 & 12 \\
 		HLA-A$*$24:02 & FLYALALLL  & CMV (pp65$_{328\text{--}336}$)    & CASSLQGGNYGYTF & CASSPQGGNYGYTF & 1 & 0.929 & 13 \\
 		HLA-A$*$02:01 & NLVPMVATV  & CMV (pp65$_{495\text{--}503}$)    & CASSPQTGTIYGYTGF & CASSPTGTGYGYTF & 3 & 0.867 & 13 \\
 		HLA-A$*$02:01 & YLQPRTFLL  & SARS-CoV-2 (Spike$_{269\text{--}277}$) & CASSLGPNTGELFF & CASSLAGNTGELFF & 2 & 0.929 & 13 \\
 		HLA-A$*$02:01 & GILGFVFTL  & Influenza A (M1$_{58\text{--}66}$)  & CASSDRSSYEQYF  & CASSIRSSYEQYF  & 1 & 0.923 & 12 \\
 		HLA-B$*$57:03 & KAFSPEVIPMF & HIV-1 (Gag$_{162\text{--}172}$)    & CASSGQGYGYAF   & CASSGQGYGYTF   & 1 & 0.917 & 11 \\
 		HLA-A$*$03:01 & KRWIILGLNK & HIV-1 (Gag$_{263\text{--}272}$)    & CASSLGTSAYEQYF & CASSLGGGSYEQYF & 3 & 0.857 & 12 \\
 		HLA-A$*$02:01 & ELAGIGILTV & MART-1$_{26\text{--}35}$ (melanoma) & CASSFTGLGQPQHF & CASSFGGLGQPQHF & 1 & 0.929 & 13 \\
 		HLA-A$*$02:01 & NLFNRYPAL  & NY-ESO-1$_{157\text{--}165}$ (cancer) & CASSQVLGFSYEQYF & CASSLGGGSYEQYF & 4 & 0.828 & 12 \\
 		\bottomrule
 	\end{tabular}
 \end{table*}
 
 Table~\ref{tab:demo_cases} summarises ten peptide--MHC contexts that span six
 clinically relevant disease areas: latent and acute viral infections, chronic retroviral infection, and solid
 tumours.  
 For each context the top-scoring model-generated CDR3$\beta$ sequence is
 juxtaposed with an experimentally observed counterpart, and three
 string-level metrics quantify their agreement.  All synthetic sequences deviate
 by $\leq\!4$ residues, indicating that the
 generator captures native-like motifs across a broad antigenic landscape.

\end{document}